\documentclass[reprint,amsmath,amssymb,aps,prl]{revtex4-1}

\usepackage{graphicx}
\usepackage{subfigure}
\usepackage{amsmath}
\usepackage{amsthm}
\usepackage{amssymb}
\usepackage{xcolor}
\usepackage{hyperref}
\hypersetup{colorlinks=true, linkcolor=black, citecolor=black, urlcolor=blue}
\usepackage{soul}

\newcommand{\ket}[1]{\left| #1 \right>} 
\newcommand{\bra}[1]{\left< #1 \right|} 

\begin{document}

\title{Open Quantum-System Simulation of Faraday's Induction Law via Dynamical Instabilities}

\date{\today}

\author{Elvia Colella}
\email{elvia.colella@uibk.ac.at}
\author{Arkadiusz Kosior}
\author{Farokh Mivehvar}
\author{Helmut Ritsch}
\affiliation{Institut f\"ur Theoretische Physik, Universit{\"a}t Innsbruck, A-6020~Innsbruck, Austria}

\begin{abstract}
We propose a novel type of a Bose-Hubbard ladder model based on an open quantum-gas--cavity-QED setup to study the physics of dynamical gauge potentials. Atomic tunneling along opposite directions in the two legs of the ladder is mediated by photon scattering from transverse pump lasers to two distinct cavity modes. The resulting interplay between cavity photon dissipation and the optomechanical atomic back-action then induces an average-density-dependent dynamical gauge field. The dissipation-stabilized steady-state atomic motion along the legs of the ladder leads either to a pure chiral current, screening the induced dynamical magnetic field as in the Meissner effect, or generates simultaneously chiral and particle currents. For sufficiently strong pump the system enters into a dynamically unstable regime exhibiting limit-cycle and period-doubled oscillations. Intriguingly, an electromotive force is induced in this dynamical regime as expected from an interpretation based on Faraday's law of induction for the time-dependent synthetic magnetic flux. 
\end{abstract}
\maketitle 

\emph{Introduction.}---Gauge theories describe a plethora of fundamental phenomena, from electromagnetism to the interaction between elementary particles~\cite{wiese2013ultracold}. Exploring limits of such theories has driven the interest toward the experimental implementation of ``synthetic'' gauge potentials in ultracold neutral atoms~\cite{dalibard2011colloquium}. Early experiment successfully realized \emph{static} background artificial gauge potentials~\cite{engels2003observation,Lin2009Synthetic,Lin2011Spin,aidelburger2011experimental,struck2012tunable,aidelsburger2013realization,aidelsburger2014measuring}; however, in a genuine gauge theory the gauge potentials appear as \emph{dynamical} degrees of freedom. In this respect, introducing a density dependence on the synthetic vector potential via the periodical modulation of two-body interactions~\cite{greschner2014density, clark2018observation} and Floquet lattice shaking~\cite{itin2015effective,gorg2018realisation} constitutes an important step towards implementing dynamical gauge potentials~\cite{goldman2014light}.

Among alternative approaches to induce dynamical gauge potentials, quantum-gas--cavity-QED setups stand out owing to the intrinsic dynamical nature of cavity fields~\cite{ritsch2013cold,Mivehvar2021Cavity}. Many interesting phenomena have been predicted to arise in systems with cavity-induced dynamical gauge potential, from the dynamical appearance of a vector potential at the onset of superradiance~\cite{Kollath2016Ultracold,colella2019Hofstadter} to a dissipation-induced dynamical Peierls phase~\cite{cooper2016superradiance} and the Meissner-like expulsion of a magnetic field~\cite{ballantine2017meissner}. The prediction of cavity-induced dynamic spin-orbit coupling~\cite{mivehvar2014synthetic,dong2014cavity,deng2014bose,mivehvar2015enhanced,Ostermann2019Cavity,Mivehvar2019Cavity,halati2019cavity,Ostermann2021Many} and its recent realization~\cite{kroeze2019dynamical} has opened a new avenue for engineering dynamical gauge potentials alongside free-space schemes and experiments~\cite{lienhard2020realization}.

\begin{figure}[b!]
\centering
\includegraphics[width=0.45\textwidth]{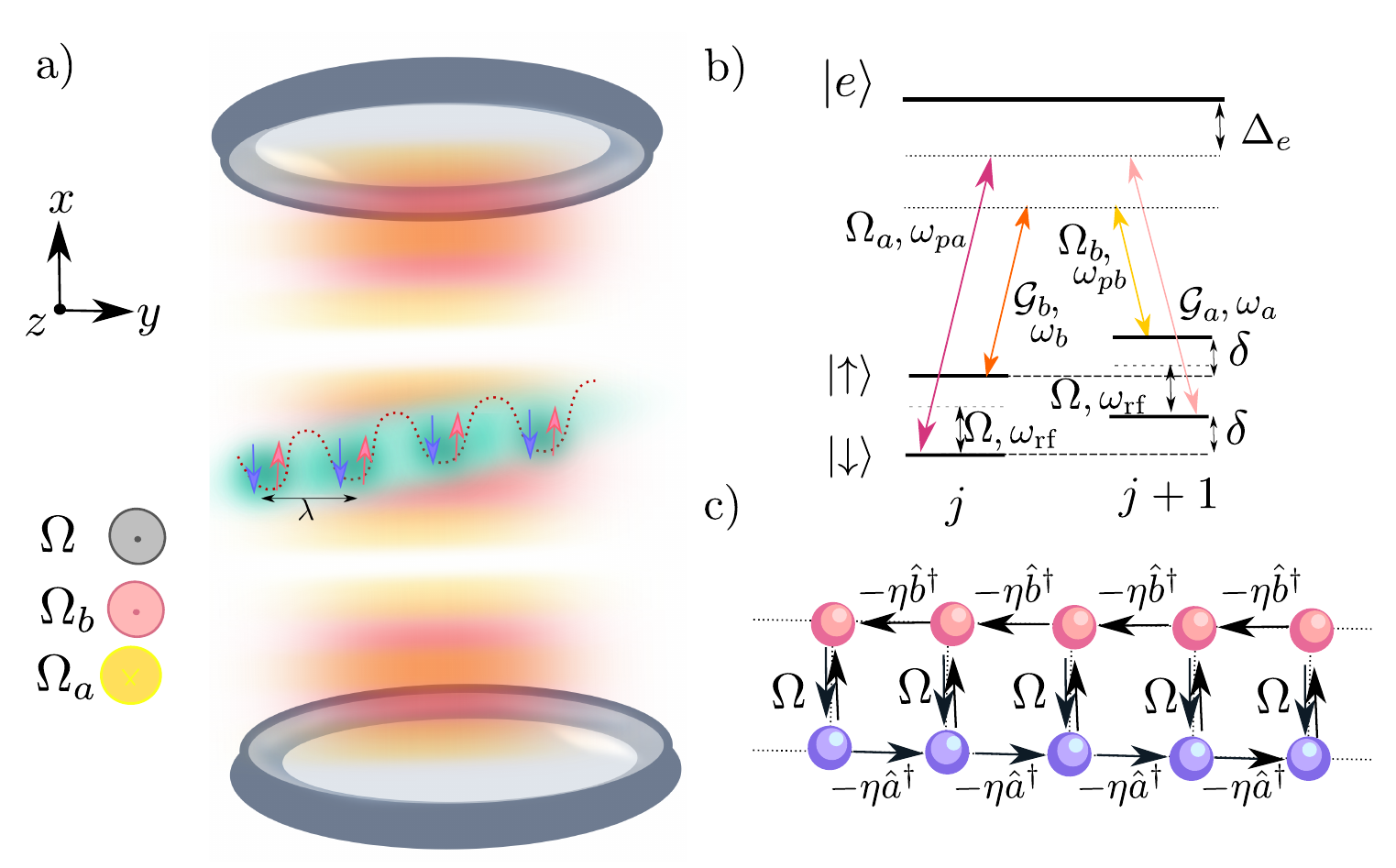}
\caption{Sketch of the setup.
(a) A spinor BEC is loaded into a 1D tilted optical lattice perpendicular to the axis of a linear cavity. Neighbouring sites are Raman coupled via two cavity modes with strengths $\mathcal{G}_{a,b}$ and transversely applied laser fields with amplitudes $\Omega_{a,b}$. A microwave couples the two atomic states locally with strength $\Omega$. (b) Sketch of the atomic level structure and of the two independent two-photon Raman transitions inducing directional tunneling between neighboring lattice sites. (c) Effective mapping of the spinor BEC in the 1D lattice in panel~(a) into a two-leg Bose-Hubbard ladder with cavity-assisted longitudinal hoppings and microwave-generated transverse tunneling.}
\label{model}
\end{figure}

Motivated by the recent experimental realization of the dynamical spin-orbit coupling~\cite{kroeze2019dynamical} and a two-mode Dicke model~\cite{morales2019two}, we propose a novel cavity-QED scheme for implementing an \emph{average-density-dependent} dynamical gauge potential. In particular, we develop a ladder model~\cite{orignac2001meissner} with cavity-assisted counterpropagating tunnelings as shown in Fig.~\ref{model}. A dynamic gauge potential appears at the onset of the superradiant photon scattering from two transverse pump lasers into two cavity modes owing to dissipation-induced phase shifts of cavity photons. In contrast to previous works~\cite{Kollath2016Ultracold,colella2019Hofstadter,cooper2016superradiance}, we take into account the  optomechanical back-action of the atomic dynamics, leading to an average-density-dependent dynamical magnetic flux when the optomechanical back-action shifts significantly the cavity resonances and consequently the dissipation-induced photonic phases. The system exhibits three steady states: Photon-balanced Meissner (PB-M) and vortex (PB-V) states and a photon-imbalanced biased ladder (PI-BL) phase. In addition, the phase diagram of the system features a region of highly nonlinear dynamics with no steady state. Two stable dynamical phases are identified with limit-cycle and period-doubled oscillations. Remarkably, an electromotive force is induced naturally in this  regime, mimicking Faraday's law of induction.  

\emph{Model.}---Consider a spinor Bose-Einstein condensate (BEC) inside a linear optical cavity. The BEC is strongly confined by an external optical lattice to a one dimension perpendicular to the cavity axis at anti-nodes of two distinct cavity electromagnetic modes, $\hat a$ and $\hat b$; see Fig.~\ref{model}(a). The natural tunneling of atoms along the lattice is suppressed by applying a potential gradient $\delta$~\cite{jaksch2003creation}. A directional hopping is restored by two independent resonant two-photon Raman transitions as shown in Fig.~\ref{model}(b): The ground pseudospin  state $\ket{\downarrow}$ ($\ket{\uparrow}$)  is coupled, respectively, by Rabi rates $\mathcal{G}_a$ ($\mathcal{G}_b$) and $\Omega_a$ ($\Omega_b$) to a far detuned excited state $\ket{e}$ via the cavity mode $\hat{a}$ ($\hat{b}$) with resonance frequency $\omega_a$ ($\omega_b$) and an out-of-plane transverse pump laser with frequency $\omega_{pa}$ ($\omega_{pb}$). The two pseudospin ground states are coupled on-site with a rate $\Omega$ through a radio-frequency drive with frequency $\omega_{\rm rf}$.

For large atomic detuning, the excited state can be adiabatically eliminated. By only retaining resonant scattering terms, the effective Hamiltonian reads~\cite{Supplementary},
\begin{align}
\hat{H}&=
-\hbar\eta\sum_{j=1}^L (\hat a^\dagger \hat c^\dagger_{\downarrow,j+1} \hat c_{\downarrow,j}
+\hat b \hat c^\dagger_{\uparrow,j+1} \hat c_{\uparrow,j}+\mathrm{H.c.})
\nonumber\\
&-\hbar\Omega \sum_{j=1}^L (\hat c^\dagger_{\uparrow,j} \hat c_{\downarrow,j}+\mathrm{H.c.}) 
\nonumber\\
&+\frac{V}{2} \sum_{j=1}^L\sum_{\sigma=\uparrow,\downarrow}\hat  N_{\sigma,j}(\hat N_{\sigma,j}-1)
+\gamma V \sum_{j=1}^L \hat{N}_{\downarrow,j}\hat{N}_{\uparrow,j}
\nonumber\\
&-\hbar(\Delta_a-U \hat N_\downarrow )\hat a ^\dagger \hat a -\hbar(\Delta_b-U \hat N_\uparrow)  \hat b^\dagger \hat b,
\label{eq:ham}
\end{align}
where $ \hat c_{\sigma,j}$ is the atomic bosonic annihilation operator for pseudospin $\sigma$ at site $j$, and $\hat{N}_{\sigma}=\sum_j \hat{N}_{\sigma,j}=\sum_j \hat c_{\sigma,j}^\dag\hat c_{\sigma,j}$. The effective model~\eqref{eq:ham} constitutes a spinor Bose-Hubbard-type Hamiltonian with cavity-induced dynamical spin-orbit coupling. It can be effectively mapped into a two-leg Bose-Hubbard ladder of length $L$, with one of the two pseudospin states acting as a synthetic dimension~\cite{celi2014synthetic}. In this spirit, the first row of the Hamiltonian corresponds to the motion of the atoms along the longitudinal direction (i.e., legs) of the ladder. The  forward tunneling amplitudes, $\hat{t}_{a(b)}\equiv\hbar \eta \hat a^\dagger(\hat b)$, are restored by scattering photons from the pump (cavity) into the cavity (pump) at a rate $\eta=\mathcal{G}_a\Omega_a/\Delta_e=\mathcal{G}_b\Omega_b/\Delta_{e}$, with $\Delta_{e}= \omega_{p}-\omega_{e}$ being the effective atomic detuning from the average pump frequency $\omega_p=( \omega_{pa}+\omega_{pb})/2$. The second line represents a transversal hopping along rungs of the ladder, with tunneling amplitude set by the radio-frequency coupling $\Omega$. The third line includes repulsive two-body on-site atomic interactions, with $V$ being the strength of intra-species interactions and $\gamma$ parameterizing the ratio between the strength of inter- and intra-species interactions. The last two terms represent the free energy of the photon fields with the cavity detunings defined as $\Delta_{a(b)}=\omega_{p}\pm\omega_{\mathrm{rf}}/2-\omega_{a(b)}$, where in the following we assume $\Delta\equiv\Delta_a=\Delta_b$. The cavity resonances are shifted by the atomic medium $U\hat{N}_\sigma$, where $U =\mathcal{G}_{a}^2/\Delta_e= \mathcal{G}_{b}^2/\Delta_{e}$ is the dispersive shift per atom.

\textit{Average-density-dependent dynamical gauge potential and synthetic magnetic field.}---Let us now describe the mechanism by which a dynamical gauge potential can arise when the photon-assisted tunnelings $\hat{t}_{a,b}$ are restored. We recall that on a lattice, the coupling of a charged particle $\mathcal{Q}$ to a vector potential $\boldsymbol{\mathcal{A}}$ is manifested by the phase factor $e^{i\mathcal{Q}/\hbar\int_{\rm{path}}\boldsymbol{\mathcal{A}}\cdot d\mathbf{l}
}$ of the tunneling amplitude, i.e., Peierls phase~\cite{peierls1933zur}. The Peierls phase is fixed by the circulation of the vector potential along the path enclosing the unit cell, and reduces to the Aharonov-Bohm phase in the continuum limit~\cite{aharonov1959significance}. Therefore, the magnetic flux piercing a lattice plaquette is
$\Phi_B =\int_{\rm{unit\, cell}}\boldsymbol{\mathcal A}\cdot d\mathbf{l}$.

For neutral atoms the vector potential must be engineered tailoring the atomic tunneling amplitudes. To achieve this, our scheme exploits the superradiant scattering of photons into the cavity. In particular, in the superradiant phase the collective synchronized emission of photons results in a macroscopic occupation of the two cavity modes, which can be treated semi-classically as dynamical electromagnetic fields. The cavity fields are thus described by coherent states~\cite{piazza2013bose}, $\hat a\rightarrow\langle \hat a \rangle\equiv\alpha=|\alpha| e^{i\phi_\alpha}$ and $\hat b\rightarrow\langle \hat b \rangle\equiv\beta= |\beta| e^{i\phi_\beta}$, and the tunneling amplitudes for the lower and upper leg respectively reduce to \emph{c-numbers} $t_a=\hbar \eta |\alpha|e^{i\phi_\alpha}$ and $t_b=\hbar \eta |\beta|e^{-i\phi_\beta}$.  The pumping geometry is equivalent to a well defined gauge choice where the transverse component of the vector potential along the rungs $\mathcal A^\perp_{j}=0$ vanishes, and its longitudinal components depend on the phases of the cavity fields $\mathcal A^\parallel_{a,j}\propto \phi_\alpha$ and $\mathcal A^\parallel_{b,j}\propto-\phi_\beta$.
Hence, the total phase acquired by the atomic wavefunction around a closed loop along one plaquette is 
$\Phi=\phi_\alpha+\phi_\beta$.
The dynamics of the atoms is equivalent to the one of charge particles $\mathcal{Q}$ threaded by the magnetic flux $\Phi_B=\hbar\Phi/\mathcal{Q}=\Phi_{0B}\Phi/2\pi$ with $\Phi_{0B}=h /\mathcal{Q}$. 

In the adiabatic limit for the photonic dynamics~\cite{Mivehvar2021Cavity,ritsch2013cold}, the cavity fields can be slaved to the atomic degrees of freedom and be obtained from the stationary solution of the Heisenberg equations of motion, $\alpha=-\eta\Theta_\downarrow /(\Delta_a-UN_\downarrow+i\kappa_a)$ and $\beta=-\eta \Theta_\uparrow^* /(\Delta_b-UN_\uparrow+i\kappa_b)$, where $\kappa_{a,b}$ are the decay rates of the cavity fields and $N_\sigma=\langle \hat{N}_\sigma \rangle$. This prescribes the phase locking between the photons and the average atomic hopping operators, $\Theta_\sigma=\sum_j\langle c^\dagger_{\sigma,j+1} c_{\sigma,j}\rangle$. Up to a global phase, the phases of the cavity fields  are uniquely determined~\cite{Supplementary},

\begin{equation}
\phi_{\alpha(\beta)}= -\arctan\left(\frac{\kappa_{a(b)}}{\Delta_{a(b)}-UN_{\downarrow(\uparrow)}}\right).
\label{phaseseq}
\end{equation}
The atomic ladder acts as a refractive medium for the light inside the resonator dispersively shifting the cavity resonances, i.e., $\Delta_{a(b)}-UN_{\downarrow(\uparrow)}$, and the magnetic flux $\Phi_B\propto\phi_\alpha+\phi_\beta$ becomes non-linearly dependent on the atomic leg density.

In order to unveil the effect of the cavity-induced magnetic field, let us consider the single-particle physics.
The atomic part of the Hamiltonian~\eqref{eq:ham} can be diagonalized to yield the single-particle atomic energy bands,  
\begin{align}
\frac{\epsilon_{\pm}(q)}{\hbar}=-\eta F_+ +\frac{U}{2}n_{\rm ph}
\pm\sqrt{ \Omega^2+\left(\eta F_- -\frac{U}{2}\Delta n_{\rm ph}\right)^2}.
\end{align} 
Here we have defined $F_\pm(q)=|\alpha|\cos(q+\phi_\alpha)\pm|\beta|\cos(q-\phi_\beta)$, $n_{\rm ph}=n_\alpha+n_\beta=|\alpha|^2+|\beta|^2$ as the total number of photons, and $\Delta n_{\rm ph}=n_\alpha-n_\beta$ as the photon-number difference. The quasi-momentum $q$ is minimally-coupled to the photon phases $\phi_{\alpha,\beta}$, and is conserved during the dynamics for the non-interacting system. The self-consistent band structure exhibits different behaviors depending on $\Delta n_{\rm ph}$. For the photon-balanced (PB) case, $\Delta n_{\rm ph}=0$, the lowest energy band has either a single minimum at $q=0$ or symmetric double minima at $q=\pm q_m\neq0$. The atomic ground state corresponds to a Meissner (M) and a vortex (V) phase, respectively. For the photon-imbalanced (PI) case, $\Delta n_{\rm ph}\neq0$, the energy bands become asymmetric, with the lowest band developing a single minimum at a nonzero quasi-momentum, $q=q_m\neq0$. The photon imbalance drives an atomic population imbalance in the two legs, which breaks the $\mathbb{Z}_2$ reflection symmetry of the system corresponding to the invariance under the exchange of the cavity modes $ \hat{a} \leftrightarrow \hat{b}^\dagger$ and ladder legs $\hat{c}_{\uparrow,j} \leftrightarrow \hat{c}_{\downarrow,j}$. We identify this state as the biased ladder (BL) phase~\cite{wei2014theory, uchino2015population, greschner2016symmetry}.

\begin{figure}[t!]
\centering
\includegraphics[width=0.5\textwidth]
{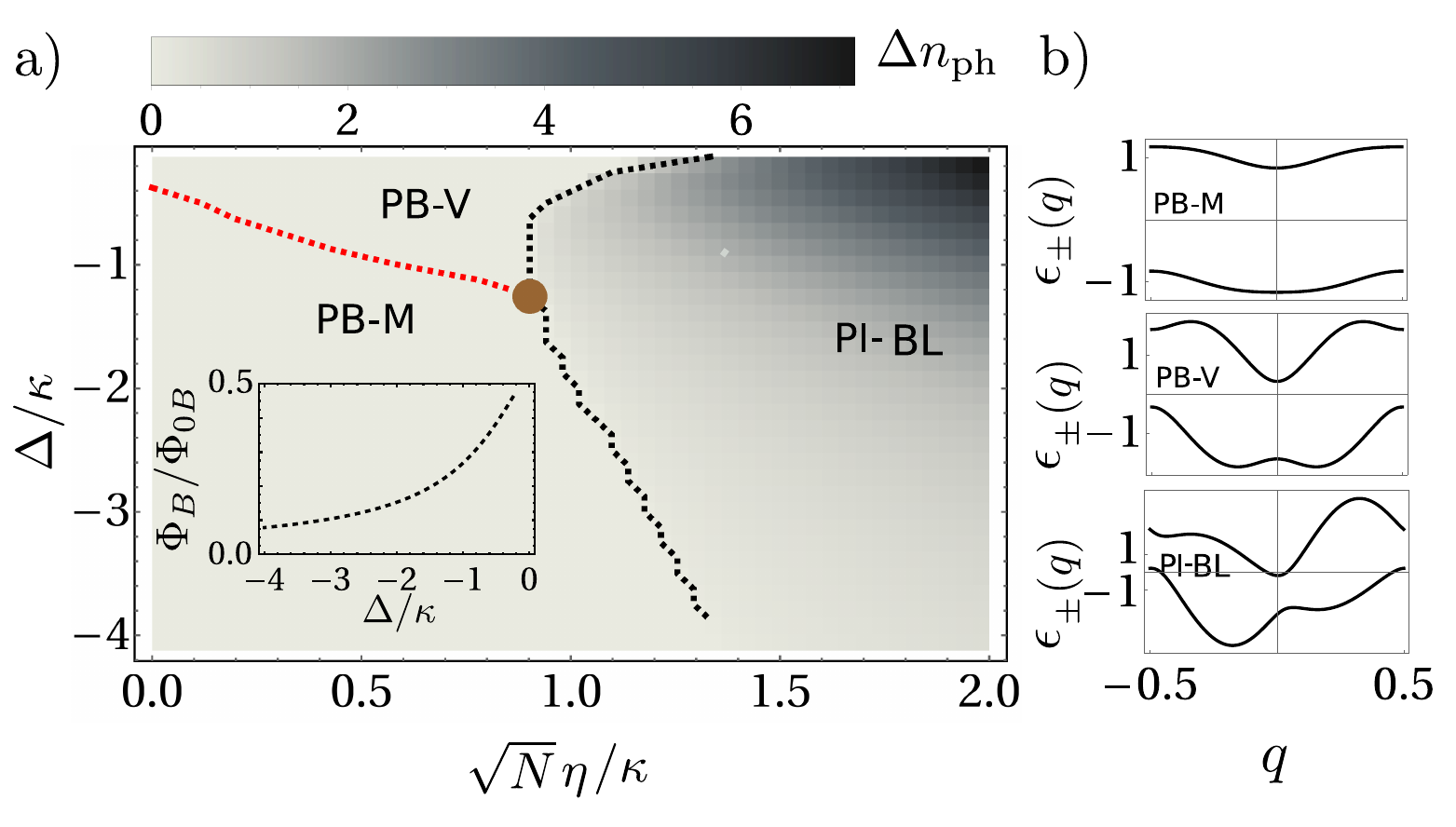}
\caption{Single-particle phase diagram. (a) The steady-state phase diagram in the $\{\sqrt{N}\eta/\kappa,\Delta/\kappa\}$ parameter plane. The self-consistent band structure exhibits three distinct phases: Photon-balanced Meissner (PB-M) and vortex (PB-V), and photon-imbalanced bias ladder (PI-BL) states. The color map indicates the photon imbalance $\Delta n_{\rm ph}$. The red dashed curve separating the PB-M and the PB-V states has been obtained analytically~\cite{Supplementary}. The three phases intersect in a tricritical point indicated by the brown dot.   
Inset: The synthetic magnetic flux $\Phi_B/\Phi_{0B}$ as a function of $\Delta/\kappa$. 
For $\Delta/\kappa=-0.5$,
(b) typical band structure $\epsilon_\pm(q)$ in each phase corresponding to $\sqrt{N}\eta/\kappa=\{0.08,0.50,1.00\}$. 
Other parameters are set to $L=51$, $U=\Delta/2L$, $\Omega=1$, and $V=\gamma=0$.}
\label{figsingleparticle}
\end{figure}

We find the steady state of the system by looking at the long-time dynamics of the coupled Heisenberg equations of motion with periodic boundary conditions, starting from a uniform density distribution with zero quasi-momentum~\cite{Supplementary}. The steady-state phase diagram in the $\Delta$--$\eta$ parameter plane is mapped out in Fig.~\ref{figsingleparticle}(a) for $N=1$.  Typical self-consistent energy bands are presented in Fig.~\ref{figsingleparticle}(b). In this non-interacting low-density regime $\bar{n}=\bar{n}_\downarrow+\bar{n}_\uparrow=1/2L\ll1$, the optomechanical back-action  $UN_\sigma$ of the atomic medium on the cavity resonances is negligible. Therefore, the magnetic flux $\Phi_B$ is almost density independent and can only be tuned by varying the cavity parameters $\Delta$ and $\kappa$; see the inset of Fig.~\ref{figsingleparticle}(a).

\begin{figure}[t!]
\centering
\includegraphics[width=0.5\textwidth]
{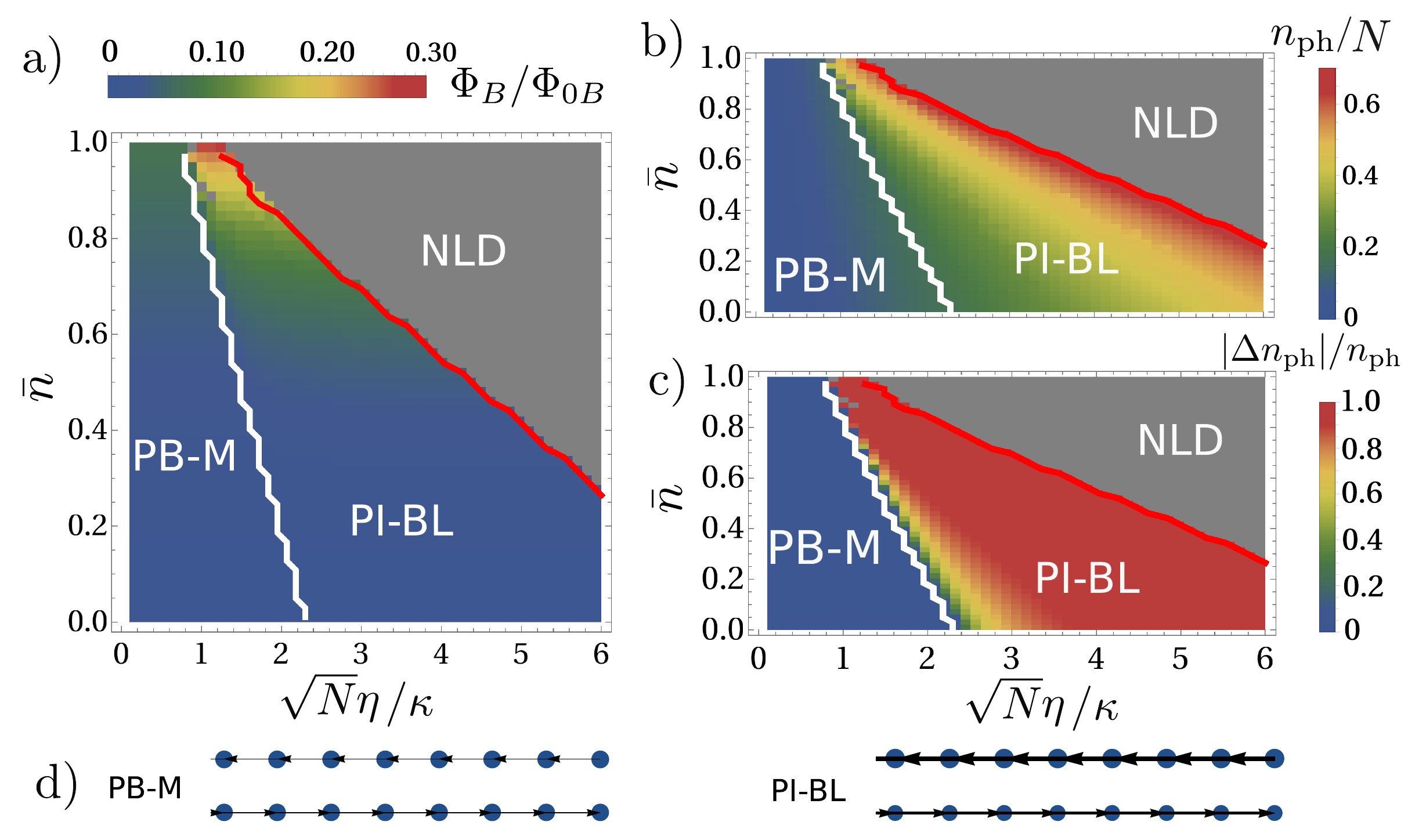}
\caption{Many-body phase diagram. (a) Non-equilibrium phase diagram in the $\{\sqrt{N}\eta/\kappa,\bar n\}$ parameter plane.
For $\Delta/\kappa=-6$, the system exhibits only two of the steady-state phases of Fig.~\ref{figsingleparticle}(a): the PB-M and the PI-BL states. The two phases are separated by the solid white line. The gray area indicates a region of dynamical instability with no steady state. 
The color map indicates the steady-state synthetic magnetic flux $\Phi_B$, clearly showing average-density dependence.
The total photon number $n_{\rm ph}$ (b) and the photon number difference $\Delta n_{\rm ph}/n_{\rm ph}$ (c) are shown in the same parameter plane. Photon number saturates the scale in the bright blue region in (b). (d) Current patterns in the PB-M and the PI-BL states. Parameters are the same as Fig.~\ref{figsingleparticle}, except $VN=1$, $\gamma=0.1$.}
\label{magneticflux}
\end{figure}

By increasing the ladder density $\bar{n}=N/2L$, the dispersive shift $UN_\sigma$ becomes significant, and the density dependence of the induced magnetic flux $\Phi_B$ becomes apparent.  Figure~\ref{magneticflux}(a) shows for weak on-site interactions the stationary value of the magnetic flux $\Phi_B/\Phi_{0B}$ as a function of the average atomic density $\bar{n}=N/2L$
and the pump strength $\sqrt{N}\eta$ for cavity detuning, $\Delta=-6\kappa$. The gray color indicates a dynamical region with no steady-state solution. Figure~\ref{magneticflux}(a) shows that density effects become relevant for higher fillings, where the two cavity modes are dispersively shifted closer to resonance. The total photon number $n_{\rm{ph}}$ and the relative photon number difference $|\Delta n_{\rm{ph}}|/n_{\rm{ph}}$ are shown in Figure~\ref{magneticflux}(b) and (c), respectively. The system exhibits a phase transition from a photon balanced to a photon imbalanced phase when the pumping strength is increased (white solid line in Figure~\ref{magneticflux}). For large cavity detunings the weak magnetic fields, $\Phi_B/\Phi_{0B}<0.15$, do not support a PB-V phase-transition at weak pumping.

\textit{Superradiance and persistent currents.}---Since each photon scattering process is accompanied by a directional atomic tunneling along the legs of the ladder, stationary currents flowing in opposite directions are generated in the superradiant phase; see the sketch in Fig.~\ref{model}(c). Dissipation plays an essential role in the generation of these currents \cite{cooper2016superradiance}. By inspection of the leg currents,
$J_\downarrow=i\eta\sum_j \langle a^\dagger \hat c^\dagger_{\downarrow,j+1} \hat c_{\downarrow,j}-\text{H.c.} \rangle$
 and  $J_\uparrow=i\eta\sum_j \langle \hat b \hat c^\dagger_{\uparrow,j+1} \hat c_{\uparrow,j}- \text{H.c.} \rangle$,
one sees that for the steady state the leg currents are proportional to the photon number of the respective coupled modes,  $J_\downarrow=2\kappa|\alpha|^2$ and $J_\uparrow=-2\kappa|\beta|^2$. The chiral current is determined by the total number of photons leaked out of the cavity, $J_{c}=J_\downarrow-J_\uparrow=2 \kappa n_{\rm ph}$, while the photon number difference identifies the net particle current, $J_{p}=J_\downarrow+J_\uparrow=2\kappa\Delta n_{\rm ph}$.
The current patterns are illustrated schematically in Fig.~\ref{magneticflux}(d). At weak pumpings, a Meissner phase is stabilized with a pair of equal, counterpropagating currents flowing along the two legs. At stronger pump strengths a net particle current is driven by the photon imbalance in the biased ladder phase~\cite{wei2014theory,uchino2015population,greschner2016symmetry}.

\begin{figure}[t!]
\centering
\includegraphics[width=0.5\textwidth]
{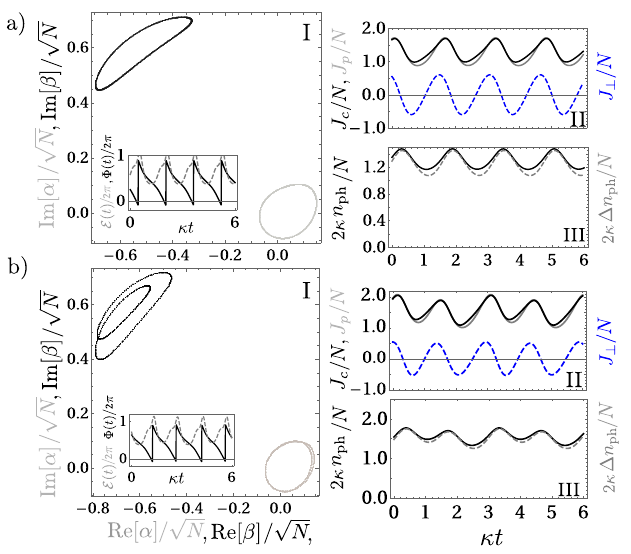}
\caption{Periodic nonlinear dynamics for high densities $\bar n=1.42$. The phase space trajectories of the two cavity-mode amplitudes $\alpha$ (gray) and $\beta$ (black) for long-time dynamics with $\sqrt{N}\eta/\kappa=1.3$ (a) and 1.55 (b) [outside of the phase diagram of Fig.~\ref{magneticflux}].
Insets in panels (I): Time evolution of the magnetic flux (black) and e.m.f (dashed gray). Time evolution of the induced chiral $J_c$ (black), particle $J_p$ (dashed gray), and rung $J_\perp$ (dotted blue) currents (II), and of the total photon number $n_{\rm ph}$ (black) and the photon number difference $\Delta n_{\rm ph}$ (dashed gray) (III). Panel (a) exhibits stable limit-cycle oscillations, while panel (b) shows period-doubled oscillations. Other parameters are the same as Fig.~\ref{magneticflux} for $N=140$.} 
\label{dianmicalunstablestate}
\end{figure}

\textit{Dynamical instabilities and Faraday's induction law.}---We now take a closer look into the grey region of Fig.~\ref{magneticflux}, where the system exhibits a highly nonlinear dynamics (NLD). When the long-time dynamics is characterized by periodic oscillation of the cavity field amplitudes [see Fig.~\ref{dianmicalunstablestate}(a) III and (b) III], the system behaves like a limit-cycle oscillator. Self-sustained periodic oscillations of the cavity modes spontaneously emerge in absence of an external periodic drive, breaking the time-translational symmetry \cite{sacha2017time}. The time-translational symmetry breaking in driven-dissipative systems has been recently interpreted as a dissipative time crystal~\cite{kessler2019emergent,Kessler2020Observation}. The system also exhibits a period-doubling bifurcation at stronger pumping with the appearance of an additional halved frequency component above the main limit-cycle oscillation frequency, possibly leading to chaos~\cite{piazza2015self}. 

The nontrivial dynamics of the photonic phases shown in Fig.~\ref{dianmicalunstablestate} leads to a time-dependent magnetic flux $\Phi_B(t)$. This in turn induces an electromotive force $\mathcal{E}(t)=-\partial \Phi_B(t)/\partial t=-(\Phi_{0B}/2\pi)\partial \Phi(t)/\partial t$, with
\begin{equation}
\frac{ \partial \Phi(t)}{\partial  t}
=2\Delta-U N-\frac{K_\downarrow}{2|\alpha|^2}-\frac{K_\uparrow}{2|\beta|^2},
\end{equation}
and the average longitudinal kinetic energies, $K_\downarrow=-2\eta\text{Re}(\alpha^*\Theta_{\downarrow})$ and $K_{\uparrow}=-2\eta\text{Re}(\beta^*\Theta_{\uparrow}^*)$.  The time evolution of the magnetic flux and the induced electromotive force are shown in the insets of Fig.~\ref{dianmicalunstablestate}(a)~I and (b)~I.
The time-dependent electromotive force drives periodically the atomic population between the two legs, apparent from the oscillating chiral $J_c$, particle $J_p$, and the emerging rung current $J_\perp=i\Omega\sum_j \langle \hat c^\dagger_{\uparrow,j} \hat c_{\downarrow,j}-\text{H.c.} \rangle$; see Fig.~\ref{dianmicalunstablestate}(a)~II and (b)~II. In contrast to the steady-state results the photon-number counts are no longer an exact measurement of the leg currents.
One can think of the residual currents as induction currents which oppose the variation of the magnetic flux, thus mimicking Faraday's law of induction with neutral particles \cite{maxwell1954a}.

\textit{Experimental considerations.}---Our proposal can be realized by driving two optical transitions of $^{87}$Rb atoms as in Ref.~\cite{kroeze2019dynamical}. Several atomic ladders can be experimentally isolated from single rows of a 2D optical lattice in the plane intercepted by the cavity axis and the $y$ direction.
The size of the ladder in the $y$ direction is strictly limited by the cavity waist $w_0$. Assuming a transversal size of $2w_0\sim 150$~$\mu$m, an atomic cloud of $\ell_y\sim 70$~$\mu$m, and an optical lattice with the lattice constant $a=\lambda_{851\mathrm{nm}}/2=0.426$~$\mu$m, the ladder would have $N_{\mathrm{u.c}}=\ell_y/a\sim 160$ unit cells. Upon redistribution of the cloud into the 2D lattice, and assuming a longitudinal cloud size of $\ell_x\sim2\ell_y$, the filling of the 2D lattice can be varied in a range of $\nu=N/(\ell_x \ell_y)\in[0.1,2]$ for an atomic cloud of $N\sim0.05-1.1\times10^5$ atoms, respectively.   The impact of the dispersive shifts $UN_\sigma$  on the cavity resonances is significant in the strong light-matter coupling limit. A good measure of the coupling strength is provided by the parameter $UN/\kappa$, expressing the ratio between the coherent and incoherent processes in the system. For a high-finesse cavity with a linewidth of $2\kappa=17$~kHz,  $U=91$~kHz  as in Ref.~\cite{elsasser2004optical}, and $N=4\times10^5$ atoms, the optomechanical back-action, $UN/\kappa \sim 2.5>1$, is in a desired range to observe the predicted phenomena. Our scheme allows to non-destructively measure the phase diagram by monitoring photons leaking out of the cavity~\cite{ritsch2013cold}.  The steady-state atomic currents can be obtained from the population count of the two cavity modes at the detector. The magnetic flux can be inferred through homodyne detection by measuring the phase of the two mode with respect to a probe laser. 

\textit{Conclusions.}---We studied the emergence of an average-density-dependent dynamical U(1) gauge potential when the motion of neutral atoms is strongly coupled to two high-$Q$ cavity modes. The gauge potential stems from the delicate interplay between the optomechanical atomic back-action on the cavity fields and photon dissipation into the environment. It differs from previously studied cavity-induced gauge potentials which do not feature any atomic-density dependence~\cite{mivehvar2014synthetic,dong2014cavity,deng2014bose,mivehvar2015enhanced,Kollath2016Ultracold,cooper2016superradiance,halati2017cavity,colella2019Hofstadter,halati2019cavity,Ostermann2019Cavity,Mivehvar2019Cavity,halati2019cavity,kroeze2019dynamical,Ostermann2021Many}.
Emerging dynamical instabilities with stable limit-cycle and period-doubled oscillations can be interpreted as driven by an effective oscillating electromotive force according to Faraday's law of induction. Our proposed scheme offers a unique possibility to explore these exotic nonequilibrium phenomena in state-of-the-art quantum-gas--cavity-QED experiments.

\begin{acknowledgments}
\textit{Acknowledgments.}---E.\,C.\ is grateful to Stefan Ostermann and Lluis Hernandez-Mula for fruitful discussions. E.\,C.\ is a recipient of a DOC Fellowship of the Austrian Academy of Sciences and acknowledges a support from the Austrian Science Fund (FWF) within the DK-ALM (W1259-N27). F.\,M.\ is supported by the Lise-Meitner Fellowship
M2438-NBL of the FWF, and the International Joint Project No.\ I3964-N27 of the FWF and the National Agency for Research (ANR) of France.
\end{acknowledgments}

%

\newpage
\widetext
\setcounter{equation}{0}
\setcounter{figure}{0}
\renewcommand{\theequation}{S\arabic{equation}}
\renewcommand{\thefigure}{S\arabic{figure}}

\widetext
\section{SUPPLEMENTARY MATERIAL}

\subsection{Effective Hamiltonian}

\textit{Internal level structure and pumping configuration.}---Consider a three-level atom $\{\bra{\downarrow},\bra{\uparrow},\bra{e}\}$ with frequencies, $\omega_\downarrow<\omega_\uparrow<\omega_e$, placed in a multi-mode linear cavity with the cavity axis oriented along the $\hat x$ direction. The atom interacts with two distinct cavity modes, $\hat a$ and $\hat b$, with frequencies $\omega_a$ and $\omega_b$, and is transversally pumped in the $\hat z$ direction by two independent lasers of frequencies $\omega_{pa}$ and $\omega_{pb}$, and intensities $\Omega_{a}$ and $\Omega_{b}$. The two atomic pseudospin ground states, $\{\bra{\downarrow},\bra{\uparrow}\}$, are coupled by a radio-frequency laser with frequency $\omega_{\mathrm{rf}}$ and intensity $\Omega$. Note that we consider the atom located at a fixed position at one of the shared maxima of intensity of the cavity fields, $\cos(k_a x_j)=\cos(k_b x_j)=1$. Hence, any space-dependence along the cavity axis due to the mode function of light fields is dropped out in the discussion below. The total Hamiltonian of the internal degrees of freedom only reads 
\begin{align}
H(t)&=\hbar \omega_{\downarrow}\ket{\downarrow}\bra{\downarrow}+\hbar \omega_{\uparrow}\ket{\uparrow}\bra{\uparrow}+\hbar \omega_{e}\ket{e}\bra{e}+\hbar\omega_a \hat a ^\dagger \hat a +\hbar \omega_b \hat b^\dagger \hat b \\
&+\left[-\hbar\Omega e^{-i\omega_{\mathrm{rf}} t}\ket{\uparrow}\bra{\downarrow}+\left(\hbar\Omega_a e^{-i\omega_{pa} t}+\hbar\mathcal{G}_a\hat a\right)\ket{e} \bra{\downarrow}+\left(\hbar\Omega_b e^{-i\omega_{pb} t}+\hbar\mathcal{G}_b \hat b\right)\ket{e} \bra{\uparrow}+\mathrm{H.c.}\right].
\end{align} 

Introducing the average pump frequency $\omega_p=( \omega_{pa}+\omega_{pb})/2$, we recast the Hamiltonian in a time-independent form by applying a unitary transformation to an appropriate co-rotating frame, 
\begin{align}
U(t)&=\exp  \Big\{i\left[-\frac{\omega_{\mathrm{rf}}}{2}\ket{\downarrow}\bra{\downarrow}+\frac{\omega_{\mathrm{rf}}}{2}\ket{\uparrow}\bra{\uparrow}+\omega_p\ket{e}\bra{e}+\left(\omega_{p}+\frac{\omega_{\mathrm{rf}}}{2}\right) \hat a^\dagger \hat a +\left(\omega_{p}-\frac{\omega_{\mathrm{rf}}}{2}\right)\hat b^\dagger \hat b\right]t\Big\}.
\label{unitary}
\end{align}
Upon the satisfaction of the resonance condition $(\omega_{pa}-\omega_{pb})=\omega_{\mathrm{rf}}$, the time-independent Hamiltonian, $\tilde H=UH(t)U^\dagger+i\hbar  (dU/dt) U^\dagger$, reads as 
\begin{align*}
\tilde H&=-\hbar \Delta_{\downarrow}\ket{\downarrow}\bra{\downarrow}-\hbar \Delta_{\uparrow}\ket{\uparrow}\bra{\uparrow}-\hbar \Delta_{e}\ket{e}\bra{e}-\hbar\Delta_a \hat a ^\dagger \hat a -\hbar \Delta_b \hat b^\dagger \hat b \\
&+\hbar\left( \Omega_a+ \mathcal{G}_a \hat a\right) \ket{e} \bra{\downarrow}+\hbar\left(\Omega_b+\mathcal{G}_b \hat b \right)\ket{e} \bra{\uparrow}  + \mathrm{H.c.}
\end{align*}
Here the atomic detunings are defined as $\Delta_{\downarrow}=-\omega_{\mathrm{rf}}/2-\omega_{\downarrow}$, $\Delta_{\uparrow}=\omega_{\mathrm{rf}}/2-\omega_{\uparrow}$,  $\Delta_{e}=\omega_p-\omega_{e}$, and the cavity detunings as $\Delta_{a}=\omega_{p}+\omega_{\mathrm{rf}}/2-\omega_{a}$ and $\Delta_{b}=\omega_{p}-\omega_{\mathrm{rf}}/2-\omega_{b}$. For large detuning $\Delta_e$, the excited state can be adiabatically eliminated and an effective Hamiltonian is obtained for the two pseudospin ground state manifold 
\begin{equation}
H_{\{\bra{\downarrow},\bra{\uparrow}\}}=H_{c}+H_{d}+H_{s}.
\end{equation}
The first term of the Hamiltonian accounts for the bare energies of the photon fields, $\Delta_{a(b)}$,
\begin{align}
H_c=&-\hbar\Delta_a \hat a ^\dagger \hat a -\hbar \Delta_b \hat b^\dagger \hat b.
\end{align}
The second term of the Hamiltonian 
\begin{align}
H_d&= \left[-\hbar\Delta_{\downarrow}+\frac{\hbar\Omega_a^2}{\Delta_e}+\frac{\hbar\mathcal{G}_a^2}{\Delta_e} \hat a^\dagger \hat a+\frac{\hbar\Omega_a g_a}{\Delta_e} (\hat a+\hat a^\dagger)\right]\ket{\downarrow}\bra{\downarrow}
&+\left[ -\hbar\Delta_{\uparrow}+\frac{\hbar\Omega_b^2}{\Delta_e}+\frac{\hbar\mathcal{G}_b^2}{\Delta_e} \hat b^\dagger \hat b+\frac{\hbar\Omega_b g_b}{\Delta_e}(\hat b+ \hat b^\dagger) \right]\ket{\uparrow}\bra{\uparrow}
\end{align}
describes the light-shifts of the two pseudo-spin energy levels induced by the interaction of the atom with the pumping lasers and the cavity modes. In particular, $\Delta_{\downarrow(\uparrow)}$ represent the detunings of the bare atomic energies in the co-rotating frame defined by Eq.\eqref{unitary}, and  $\Omega_{a(b)}^2/\Delta_e$ are the light shifts induced by the pump. The third terms arise from consecutive absorption and emission processes of cavity photons from the same mode $\sim \hbar\mathcal{G}_{a(b)}^2/\Delta_e$. The fourth terms pump the cavity by scattering photons from the pump inside the optical resonators, shifting the atomic energy levels by $\sim\mathcal{G}_{a(b)}\Omega_{a(b)}/\Delta_e$. 

The last term  of the Hamiltonian
\begin{equation}
H_s=\hbar\left[\Omega+\frac{\Omega_a \Omega_b}{\Delta_e}+\frac{\Omega_a g_b}{\Delta_e}\hat b
+\frac{\Omega_b g_a}{\Delta_e}\hat a^\dagger+\frac{g_a g_b}{\Delta_e}\hat  a^\dagger \hat b \right]\ket{\downarrow}\bra{\uparrow}+\mathrm{H.c.},
\end{equation}
includes all the spin-flip processes that can take place either because they are externally driven (as in the case of the radiofrequency $\Omega$ in the first term), or because they result from two-photon scattering processes. In particular, the second term is a classical Raman transition induced by the two pumps, the third term is a cavity-mediated Raman transition determined by exchange between the pump and cavity fields, and the fourth term is a Raman transition solely induced by the cavity fields which redistributes photons between the two modes.

From here on, for the sake of simplicity we consider  $-\Delta_{\downarrow}+\Omega_{a}^2/\Delta_e=-\Delta_{\uparrow}+\Omega_{b}^2/\Delta_e$, and use a symmetric configuration of the following parameters: $\eta\equiv \Omega_a g_a/\Delta_e= \Omega_b g_b/\Delta_e$ and $U\equiv g_a^2/\Delta_e= g_b^2/\Delta_e$.

\textit{Tight-binding model.}---Now that all the  relevant transitions are made explicit, we turn to describe the setup for the realization of a two-component Bose-Hubbard model with cavity mediated hopping. Consider an ensemble of many atoms placed in a linear optical cavity with the same configuration pumping described in the previous section, with the internal level structure adiabatically following the center of mass motion of the particles. The atoms are strongly confined to one dimension at one of the shared maxima of intensity of the cavity fields, $\cos(k_a x_j)=\cos(k_b x_j)=1$. We can thus neglect the motion along the cavity axis  ($x$-direction) and along the pump direction ($z$-direction), and focus solely on the one-dimensional motion along the $y$-direction. An accelerated optical lattice with lattice constant $\lambda$ is placed perpendicularly to the cavity axis (i.e., along $y$-direction) and strongly confines the atoms at the lattice sites. The acceleration of the optical lattice induces a constant energy gradient, $\delta$, between neighbouring sites which prevents the natural hopping of particles along the lattice direction. By carefully tuning to resonance the frequencies of the pumping lasers, it is possible to tune out of resonance the cavity mediated spin-mixing transitions and considerably simplify the model. In particular, with the resonance conditions,   $\omega_{pa}-\omega_a=\delta$ and
 $\omega_{b}-\omega_{pb}=\delta$, only the externally driven first term of Hamiltonian $H_s$ become relevant. Besides, the backward and forward tunneling along the lattice are directionally decoupled for the two pseudo-spin ground states in $H_d$. For the lower (upper) spin state only the forward (backward) directional tunneling is tuned to resonance with the condition, $\omega_{pa}-\omega_a=\delta$ ($\omega_{b}-\omega_{pb}=\delta$).
  
The Hamiltonian in the tight-binding limit reduces to 
\begin{align}
H&=-\hbar(\Delta_a-U N_{\downarrow}) \hat a ^\dagger \hat a -\hbar (\Delta_b -U N_{\uparrow})\hat b^\dagger \hat b\nonumber \\&-\eta\sum_j (\hat a^\dagger \hat c^\dagger_{\downarrow,j+1} \hat c_{\downarrow,j}+\hat b \hat c^\dagger_{\uparrow,j+1} \hat c_{\uparrow,j}+\mathrm{h.c.})-\Omega \sum_j (\hat c^\dagger_{\downarrow,j} \hat c_{\uparrow,j}+\hat c^\dagger_{\uparrow,j} \hat c_{\downarrow,j}).
\end{align}
The second internal ground states of the atoms effectively acts as a synthetic dimension, and the system can be treated as a ladder where the longitudinal tunneling is mediated by the photons, $\hat a$ and $\hat b$, and the transversal hopping is set by the radiofrequency $\Omega$.

\textit{Origin of the synthetic magnetic field.}---Consider now the total phase acquired by the atomic-wavefunction of an atom travelling on a closed trajectory along one plaquette of the ladder. The phase of the cavity photon $\hat a^\dagger$ is first imprinted on the atom by hopping between neighbouring sites of the lower leg, $\ket{j,\downarrow}$ to $\ket{j+1,\downarrow}$.  Hopping along the synthetic direction does not imprint any phase, $\ket{j+1,\downarrow}$ to $\ket{j+1,\uparrow}$. In the reverse direction on the upper leg, $\ket{j+1,\uparrow}$ to $\ket{j,\uparrow}$, the atom will acquire the phase mediated by the second cavity field, $\hat b^\dagger$. Finally, the atom comes back to the initial site without acquiring any phase,  $\ket{j,\uparrow}$ to  $\ket{j,\downarrow}$. It is clear that the total phase acquired by the wave-function in the loop depends on the phases of the two modes, $\hat a^\dagger$ and $\hat b^\dagger$. This cavity-imprinted phase is responsible for the emergence of a synthetic magnetic field piercing the ladder plaquette. 

\subsection{Derivation of Equations 3 and 4} 
\textit{Derivation of the single-particle energy bands.}---From here on, the photon fields will be treated as classical coherent states, $\alpha\equiv\langle a\rangle =|\alpha| e^{i \phi_a}$ and $\beta\equiv\langle b \rangle=|\beta| e^{i \phi_b}$.  Within this approximation the non-interacting atomic Hamiltonian can be easily diagonalized in momentum space in the thermodynamic limit, $\hat c_{\sigma,j}=\sum_q e^{i q j} \hat c_{\sigma,q}$. The Hamiltonian in momentum space reads as
\begin{equation}
H=\sum_q+\hbar U|\alpha|^2\hat c^\dagger_{\downarrow,q}\hat c_{\downarrow,q}+\hbar U|\beta|^2\hat c^\dagger_{\uparrow,q}\hat c_{\uparrow,q}- \left(\hbar\eta |\alpha| e^{-i (q+\phi_\alpha)} \hat c^\dagger_{\downarrow,q}\hat c_{\downarrow,q}+\hbar\eta |\beta| e^{-i (q-\phi_\beta)} \hat c^\dagger_{\uparrow,q}\hat c_{\uparrow,q}+\hbar\Omega\hat c^\dagger_{\downarrow,q}\hat c_{\uparrow,q}+ \mathrm{H.c.}\right),
\end{equation} 
and can be cast in diagonal form, $
H=\sum_{q=\pm} \epsilon_{\pm,q} \gamma^\dagger_{\pm,q}\gamma_{\pm,q}$, where the two energy bands are parametrised in terms of the photon amplitudes $\alpha$ and $\beta$ as
\begin{align}
\frac{\epsilon_{\pm,q}}{\hbar}&=+\frac{U}{2}\left(|\alpha|^2+|\beta|^2\right)-\eta|\alpha|\cos\left(q+\phi_\alpha\right)-\eta|\beta|\cos\left(q-\phi_\beta\right)\nonumber\\&\pm\sqrt{\Omega^2+\left[\eta\left(|\alpha|\cos\left(q+\phi_\alpha\right)-|\beta|\cos\left(q-\phi_\beta\right)\right)-\frac{U}{2}\left(|\alpha|^2-|\beta|^2\right)\right]^2}.
\end{align}

By inspection of the band structure, it can be noted that the phases of the cavity fields, $\phi_a$ and $\phi_b$, couple to the atomic momentum  as a vector potential and relatively shift the minimum of the original uncoupled tight-binding bands by the value of the magnetic flux piercing one plaquette,  $\Phi_B/\Phi_{0B}=\phi_a+\phi_b$, with $\Phi_{0B}=\hbar/\mathcal{Q}$ being the mangnetic flux quantum and $\mathcal{Q}$ the synthetic charge of atom.  The presence of a transversal hopping along the ladder rungs hybridizes the two bands, opens a gap and shifts the minima of the band structure. For $|\alpha|=|\beta|=\sqrt{n_{\rm{ph}}/2}$, in analogy to spin-orbit-coupled BEC the band structure is characterized by a single or double minima, which for a ladder correspond, respectively, to the Meissner and vortex phase. The transition point from the Meissner to the vortex phase is thus expected at the splitting of the single minima of the band structure into two-degenerate minima. Assuming the magnetic flux $\Phi_B/\Phi_{0B}=\phi_a+\phi_b$ is piercing a plaquette of the ladder, the well known threshold from the Meissner to the vortex transition can be cast in terms of the photon number, the pump strength and the magnetic flux as, 
\begin{equation}
\eta_{c}=\frac{\sqrt{2}\Omega}{\left[\sqrt{n_{ph}}\sin(\frac{\Phi_B}{2\Phi_{0B}})\tan({\frac{\Phi_B}{2\Phi_{0B}}})\right]_{\eta=\eta_c}}.
\end{equation}
In our case this is a nonlinear, non-analytical equation, as the photon number intrinsically depends on the pump strength $n_{\rm{ph}}=n_{\rm{ph}}(\eta)$. In order to obtain the transition threshold shown in the main text as a dashed red line in Fig.~\ref{figsingleparticle}a, we have compared our numerical results for the effective hopping, $J_{\rm{eff}}=\eta n_{\rm{ph}}(\eta)$ with the critical hopping $J_{c}=\eta_c \sqrt{n_{\rm{ph}}(\eta_c)}$ -- note that the product $\eta_c \sqrt{n_{\rm{ph}}(\eta_c)}$ is independent of the pump strength and is thus well defined --. The critical threshold is then obtained as the first numerical point for which the tunneling $J_{\rm{eff}}$ exceed the critical hopping $J>J_{crit}$. The critical pump strength $\eta_c$ can be then readily extracted back.

\textit{Phase locking.}---We now derive the steady state value of the photonic phases $\phi_{a(b)}$, from which we show that the induced magnetic flux $\Phi=\phi_a+\phi_b$ acquires a dynamical dependence on the atomic occupation of the two legs. In the adiabatic approximation for cavity field dynamics, the photonic degrees of freedom can be cast in terms of the atomic operators by imposing the steady-state condition on the equations of motion for the cavity amplitudes 
\begin{align}
i \frac{\partial \alpha}{\partial t} &=-(\Delta+i\kappa-U N_\downarrow )\alpha-\eta\Theta_{\downarrow}=0,\label{eqst1}\\
i \frac{\partial \beta}{\partial t} &=-(\Delta+i\kappa-U N_\uparrow )\beta-\eta\Theta^{*}_{\uparrow}=0\label{eqst2},
\end{align}
where we have defined $\Theta_{\sigma}=\sum_i\langle c^\dagger_{\sigma,i+1}c_{\sigma,i}\rangle=|\Theta_\sigma|e^{i \phi_\sigma}$ as the average of the spin resolved hopping operator.
By only considering the steady-state solution of these equations,
\begin{align}
|\alpha|&=\frac{\eta}{\sqrt{\left(\Delta-UN_\downarrow\right)^2+\kappa^2}}|\Theta_{\downarrow}|,\\
\phi_a&=-\arctan\left(\frac{\kappa}{\Delta-U N_\downarrow}\right)+\phi_\downarrow,\\
|\beta|&=\frac{\eta}{\sqrt{\left(\Delta-UN_\uparrow\right)^2+\kappa^2}}|\Theta_{\uparrow}|,\\
\phi_b&=-\arctan\left(\frac{\kappa}{\Delta-U N_\uparrow}\right)-\phi_\uparrow,
\end{align}
we observe that the phase of the photons $\phi_{a(b)}$ and the phase of the hopping operator $\phi_{\downarrow(\uparrow)}$ are mutually dependent.  Due to momentum conservation the initial state population in momentum state is conserved. Then if particles are equally distributed among the legs and  prepared in the same momentum state $k_0$  at $t=0$, the phase of the translation operator, $\Theta_\sigma(t=0)=N/2e^{ik_0}$ exactly coincides with the initial momentum state $k_0$ and will not evolve in time due to the momentum conservation of the non-interacting system.  

If we use a uniform momentum distribution, $k_0=0$, the stationary phases $\phi_{a(b)}$ are then fixed as we have reported in the main text
\begin{align}
\phi_a&=-\arctan\left(\frac{\kappa_a}{\Delta_a-U N_\downarrow}\right),\\
\phi_b&= - \arctan\left(\frac{\kappa_b}{\Delta_b-U N_{\uparrow}}\right).
\end{align} 
Note that these phases do not evolve arbitrarily but are fixed by the dissipation constants and the detuning of the cavity fields. The effective detuning of the two cavity modes is shifted by the number of atoms in each leg, which self-consistently adapts to the state of the system. The induced effective gauge potential is therefore entirely dynamical.

\section{Weakly interacting regime}
If we focus on the weakly interacting regime of two-body repulsive interactions, the many-body atomic wave-function can be approximated as a product state of single particle wave-functions.  We can therefore substitute the atomic operators with their mean-field amplitude  $\hat c_{i,\sigma}\leftrightarrow \langle \hat c_{i,\sigma}\rangle=\psi_{i,\sigma}$, which satisfy the normalization condition $\sum_{i,\sigma}|\psi_{i,\sigma}|^2=N$. The dynamics of the system is then described by a set of non-linear equations coupled to the Heisenberg equations of motions for the cavity fields,
\begin{align}
i \frac{\partial  \psi_{i,\downarrow}}{\partial t}&=U|\alpha|^2\psi_{i,\downarrow}-\eta \alpha \psi_{i+1,\downarrow}-\eta \alpha^* \psi_{i-1,\downarrow} -\Omega \psi_{i\uparrow}\nonumber+\frac{V}{\hbar}|\psi_{i,\downarrow}|^2\psi_{i,\downarrow}+\frac{\gamma V}{\hbar} \nonumber|\psi_{i,\uparrow}|^2\psi_{i,\downarrow},\nonumber\\
i \frac{\partial \psi_{i,\uparrow}}{\partial t} &=U|\beta|^2\psi_{i,\uparrow}-\eta( \beta^* \psi_{i+1,\uparrow}+\beta \psi_{i-1,\uparrow}) -\Omega \psi_{i\downarrow}\nonumber+\frac{V}{\hbar}|\psi_{i,\uparrow}|^2\psi_{i,\uparrow}+\frac{\gamma V}{\hbar} \nonumber|\psi_{i,\downarrow}|^2\psi_{i,\uparrow},\\
i \frac{\partial \alpha}{\partial t} &=-(\Delta+i\kappa-U N_\downarrow)\alpha-\eta\sum_i \psi_{i+1,\downarrow}^*\psi_{i,\downarrow},\nonumber\\i \frac{\partial \beta}{\partial t} &=-(\Delta+i\kappa-U N_\uparrow)\beta-\eta\sum_i \psi_{i,\uparrow}^*\psi_{i+1,\uparrow}.
\end{align}
 In the main text we solved these equations for $V=1$ and $\gamma=0.1$ and looked at the long-time dynamics of the atomic and photonic states. As an initial condition we used a uniform density distribution with an equal number of particles distributed in the two legs for the atomic wave-function, and a random initial seed for the photon amplitude $\alpha$ and $\beta$. Interspecies contact interactions do not qualitatively change the physics of the system but rather push to higher energies the transition threshold for the photon-imbalanced state and for the onset of the nonlinear dynamical regime. In contrast intraspecies interactions do not affect the transition to the nonlinear dynamical regime, but  affect the transition threshold to the photon imbalanced regime.

\section{Derivation of Equation 5} 
We provide here additional insight on the dynamics of the system at strong pump and for high densities, where the system fails to reach a stationary state. In this region  the adiabatic approximation for the cavity modes breaks down and the full time dynamics of the coupled Heisenberg equations of motion must be taken into account. The Heisenberg equations of motion are a set of  complex equations and can be cast in terms of the amplitudes $|\alpha|,|\beta|$ and the phases $\phi_a,\phi_b$ of the cavity fields:  
\begin{align}
\frac{\partial |\alpha|}{\partial t} =-\kappa|\alpha|+\frac{J_\downarrow}{2|\alpha|}, &\qquad\qquad  \frac{\partial |\beta|}{\partial t} =-\kappa|\beta|-\frac{J_\uparrow}{2|\beta|}; \\
\frac{\partial \phi_a}{\partial t} =(\Delta-U N_\downarrow)-\frac{K_\downarrow}{2|\alpha|^2},&\qquad\qquad
\frac{\partial \phi_b}{\partial t} =(\Delta-U N_\uparrow)-\frac{K_\uparrow}{2|\beta|^2}.
\end{align}
 Above we have defined the mean-field average longitudinal kinetic energy for the upper leg,  $K_\uparrow=-2\mathrm{Re}(\eta\beta^*\Theta_{\uparrow}^*)$, and lower leg, $K_\downarrow=-2\mathrm{Re}(\eta\alpha^*\Theta_{\downarrow})$,  and the mean-field average longitudinal currents on the upper leg, $J_\uparrow=2\eta\mathrm{Im}(\Theta_{\uparrow}^*\beta^*)$, and lower leg, $J_\downarrow=-2\eta\mathrm{Im}(\Theta_{\downarrow}\alpha^*)$, obtained from the current operators defined in the main text. 

The equations of motion for the total number of photon, the photon number difference and the magnetic flux piercing the plaquette can be derived as follows,
\begin{align}
\frac{\partial n_{\rm{ph}}}{\partial t} &=-2\kappa n_{\rm{ph}}+J_\downarrow-J_\uparrow,\label{eqnph}\\
\frac{\partial \Delta n_{\rm{ph}} }{\partial t} &=-2\kappa \Delta n_{\rm{ph}} +J_\downarrow+J_\uparrow,\label{eqndph}\\
\frac{\partial \Phi}{\partial t} &=2\Delta-U N-\frac{n_{\rm{ph}} \left(K_\uparrow+K_\downarrow\right)+\Delta n_{\rm{ph}}\left(K_\uparrow-K_\downarrow\right)}{n_{\rm{ph}}^2-\Delta n_{\rm{ph}}^2}.
\label{eqphi}
\end{align}
From inspection of the Eqs.~\eqref{eqnph} and~\eqref{eqndph} it can be directly seen that the chiral current $J_c=J_\downarrow-J_\uparrow$ and the particle current $J_p=J_\uparrow+J_\downarrow$ act as a source term for the photon number and the photon number difference. The dissipation constant $\kappa$ is responsible for the decay of their amplitude inside the cavity, as expected. The non-trivial dynamics of the photon number induces a time-dependent leg current and a time-dependent magnetic flux according to Eq.~\eqref{eqphi}. Such phenomena is therefore reminiscent of the behaviour of charged particles in a time-dependent magnetic field where the appearance of an electromotive force (e.m.f), $\mathcal{E}=-\partial \Phi_B(t)/\partial t$, opposes to the time variation of the magnetic flux, $\Phi_B(t)$. The self-emerging electric field generating the e.m.f is oriented along the lattice direction $\hat y$ and is given for each leg by the variation of the phase of the corresponding coupled photon field,
\begin{equation}
\frac{\mathcal{Q}}{\hbar}\boldsymbol{\mathcal E}=-\frac{\mathcal{Q}}{\hbar}\frac{\partial\boldsymbol{\mathcal A}}{\partial t}=-\begin{pmatrix}
\dot{\phi_b} \\
\dot{ \phi_a} \\
\end{pmatrix}\hat e_y,
\end{equation}
with $\mathbf{\mathcal{A}}$ being the vector potential on links between longitudinal lattice sites.

\section{Mesoscopic ladders}

\begin{figure}[t]
\centering
\includegraphics[width=1\textwidth] {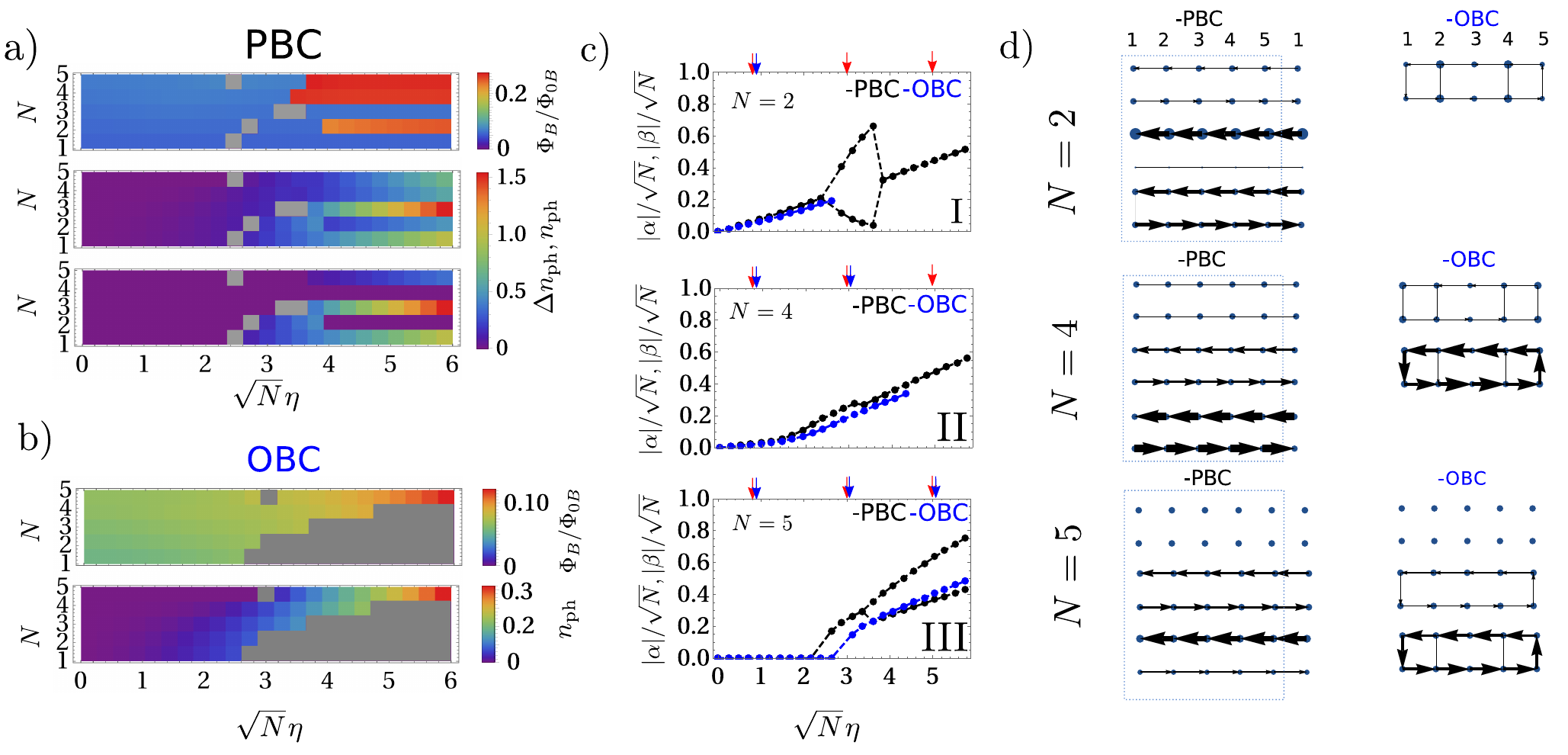}
\caption{(a) Converged magnetic flux, total photon number and photon number difference from top to bottom panel, for a ladder of length $L=5$ and increasing number of particles $N=1,2,3,4,5$. (PBC). (b) Converged magnetic flux and total photon number from top to bottom for a ladder of length $L=5$ (OBC). The photon number difference is not reported for OBC as it vanishes identically in the parameter space. In gray regions of dynamical instability for (a) and (b). (c) Direct comparison of the rescaled photon amplitude of the two modes, $|\alpha|/\sqrt{N}$ and $|\beta|/\sqrt{N}$, for PBC (black) and OBC (blue) for $N=2,4,5$ in I, II and III respectively. The red (blue) arrows on top of each panel show the pumping strengths for which the PBC (OBC) currents are  obtained in (d). Other parameters are same as Fig.~\ref{magneticflux} of the main text.}
\label{mesoscopicrings}
\end{figure}  

In this section, we solve the few-body problem with a self-consistent exact diagonalization method (SC-ED). We assume that the entanglement between atoms and photons is negligible, and treat the photonic degrees of freedom semiclassically, $\hat a\rightarrow \langle \hat a \rangle \rightarrow \alpha$ and $\hat b\rightarrow \langle \hat b \rangle \rightarrow \beta$. The atomic Hamiltonian $H_{\alpha,\beta}$ is parametrized in terms of the cavity amplitudes and is exactly solved, giving full access to the effects of correlations induced by two-body interactions. Throughout the section we will work with ladders of finite size $L$ and with conserved number of particles $N$.

\textit{Mesoscopic ladders.}---We now briefly comment on the role of density correlations for weak pumps, where the physics is dominated by onsite interactions. We study the few particle physics in mesoscopic ladders using a self-consistent exact diagonalization method with periodic boundary conditions. Notably, two-particle correlations give rise to a pump-strength threshold at half-filling for the onset of superradiance and appearance of the gauge potential. 
We identify a transition between a PB-M and a PI-BL state by increasing the pump strength, in agreement with the mean-field results. 

For open boundary conditions on the other hand, the current conservation forces the two cavity modes to be populated equally, thus hindering the transition to the photon-imbalanced regime. However, in contrast to Ref.~\cite{cooper2016superradiance}, the system is characterized by a steady-state superradiant phase with loop current of the size of the whole system. 

\paragraph{Algorithm.---} The SC-ED algorithm searches self-consistent solutions of the atomic ground state by optimizing the values of the light field amplitudes, $\alpha$ and $\beta$, which act as variational parameters. The matrix elements of the atomic Hamiltonian, $\mathcal{H}_{nm}(\alpha,\beta) = \bra{n}H_{\alpha,\beta}\ket{m} $, are parametrized in terms of the cavity amplitudes, and written in the many-body basis of Fock states  ${\{ \ket{n}\} = \ket{n_{1\downarrow},n_{2\downarrow},...,n_{L\downarrow},n_{1\uparrow},n_{2\uparrow},...,n_{L\uparrow}}}$. Here, each $n_{i\sigma}$ represents the occupation of the lattice site, $i$, on the upper ($\sigma=\uparrow$) or lower ($\sigma=\downarrow$) leg of the ladder. The algorithm is based on two fundamental steps which are repeated until convergence: diagonalization and calculation of the variational parameters $\alpha$ and $\beta$.

The algoritm starts with the initialization of the cavity amplitudes to a random guess,  $\alpha_0$ and $\beta_0$. The diagonalization step determines the lowest energy many-body state,$\Psi_{GS}(\alpha_0,\beta_0)$, which corresponds to the atomic Hamiltonian, $H_{\alpha_0,\beta_0}$. As a second step, the cavity fields $\alpha_{\rm{new}}$ and $\beta_{\rm{new}}$ are up-dated according to the stationary equations ~\eqref{eqst1} and~\eqref{eqst2}, which are averaged over the ground state, $\Psi_{GS}(\alpha_0,\beta_0)$. The convergence criteria measures the distance of the new solution to the old one, $c=1/2\sqrt{(\alpha_{\rm{new}}-\alpha_0)^2+(\beta_{\rm{new}}-\beta_0)^2}$. If the criteria is less than a certain tolerance $t=10^{-7}$, the algorithm output is given by the converged values of the cavity amplitudes and atomic ground state. If the criteria exceeds the tolerance, new matrix elements $\mathcal{H}_{nm}(\alpha_{\rm{new}},\beta_{\rm{new}})$ are calculated based on the new averages of the cavity amplitudes, $\alpha_{\rm{new}}$ and $\beta_{\rm{new}}$. The procedure is repeated until convergence. 

\paragraph{Phase diagram.---} In Fig.~\ref{mesoscopicrings} we report results for a mesoscopic ladder of size $L=5$ for both periodic (PBC) and open boundary conditions (OBC). The phase diagram is shown in Fig.~\ref{mesoscopicrings}(a) for PBC and in  Fig.~\ref{mesoscopicrings}(b) for OBC for increasing pumping strength $\sqrt{N}\eta$ and number of particles $1<N<5$. Note that, given the small system size and discrete number of particles, the phase diagram obtained in this section can only be compared to longitudinal cuts of Fig.~\ref{magneticflux} in the main text corresponding to densities $n=\{0.1,0.2,0.3,0.4,0.5\}$.    
Overall we find good qualitative  agreement between the mean field results obtained via time evolution of the Schrodinger equations and the behaviour of the system predicted by SC-ED. In the top panels of Fig.~\ref{mesoscopicrings}(a) and (b), the converged magnetic flux $\Phi_B/(\Phi_{0B}\pi)$ shows a prominent density dependence in agreement with the mean-field results. In the middle panels, the total photon number $n_{\rm{ph}}$ gradually increases confirming the absence of a superradiant threshold with the exception of $N=5$. This is shown in Fig.~\ref{mesoscopicrings}(c) where the photon amplitudes $|\alpha|/\sqrt{N}$ and $|\beta|/\sqrt{N}$ for $N=5$ are reported in full (bottom panel). The strong suppression of the photon amplitude at half-filling implies a reduction of the kinetic energy along the ladder direction, thus characterizing a Mott insulating state for the atoms. The bottom panel in Fig.~\ref{mesoscopicrings}(a) shows the photon number difference $\Delta n_{\rm{ph}}$ for PBC. We identify a transition from a photon balanced regime ($\Delta n_{\rm{ph}}=0$) at weak pumping to a photon imbalanced regime ($\Delta n_{\rm{ph}}\neq0$) at strong pumping. The two regimes are characterized by the current patterns shown in Fig.~\ref{mesoscopicrings}(d), and correspond to the Meissner and biased ladder current pattern also discussed in the main text. Results of $\Delta n_{\rm{ph}}$ for OBC are not reported in  Fig.~\ref{mesoscopicrings}(b) because the photon number difference vanishes identically. With OBC the boundary enforces current conservation between the two legs, resulting in an equal photon number for the two modes and loop currents of the size of the whole system Fig.~\ref{mesoscopicrings}(d).

\paragraph{Finite size effect.---} We come now to describe some  effects that arise due to the finite size of the system. We identify an unusual behavior of the converged ground state for commensurate filling $N=2$ and $N=4$: anomalous photon-balanced states are stabilized and exhibit counter-propagating currents similarly to the Meissner phase [Fig.~\ref{mesoscopicrings}(d)]. For incummensurate filling $N=1$, $N=3$ and $N=5$, the usual photon-balanced vs photon-imbalanced transition is retrieved instead. The unusual behaviour distinguishing odd (incommensurate) to even (commensurate) filling could depend on the finite size of the system and the survival of such phases is thus not guaranteed in the thermodynamic limit.

\end{document}